# Synergy Between Excluded Volume Effect with Co-embedded Microparticles and Chemical Doping in Carbon Nanotube Network-based Composites to Enhance Thermoelectric Power Factor

Oluwasegun Isaac Akinboye[1], Yu Zhang[1], Vamsi Krishna Reddy Kondapalli[1], Lars Alexander Olivan[4], Nilesh Raut[1], Mark Jackson[2], Adekoya Oluwaseun[1], Ryan White[2,4], Vesselin Shanov[1,3] and Je-Hyeong Bahk[1,2]

[1]Department of Mechanical and Materials Engineering, University of Cincinnati, Cincinnati, Ohio 45221, United States

[2]Department of Electrical Engineering and Computer Science, University of Cincinnati, Cincinnati, Ohio 45221, United States

[3]Department of Chemical and Environmental Engineering, University of Cincinnati, OH, 45221, United States

[4]Department of Chemistry, University of Cincinnati, OH, 45221, United States

**ABSTRACT**

*There is a growing momentum in recent thermoelectric materials research for flexible materials which enhance power output and efficiency at human wearable temperatures. In our previous work, we established the method to improve thermoelectric properties with co-embedding microparticles in carbon nanotube (CNT) network-based composites. In this work, we investigate the synergy of excluded volume effect by silica microparticles with CNT doping toward the enhancement of thermoelectric power factor. We find that the power factor can be enhanced to 28.46 μW/mK$^2$ when the CNT is doped with 0.5 μg/ml of 1,5,7-triazabicyclo [4.4.0] dec-5-ene (TBD) before co-embedding silica microparticles within the carbon nanotube-based composite. We also fabricated and demonstrated flexible wearable thermoelectric devices made of the*

*developed composites that exhibits a power output of 0.025 µW/cm² at a temperature differential of only 4 K.*

**KEYWORDS:** carbon nanotube networks, thermoelectric transport, silica microparticles, doping, flexible thermoelectric material

## Introduction

Harnessing the properties of specialized materials, thermoelectric energy conversion provides an excellent opportunity to use the heat from one's body to power wearable devices, with clean, solid-state energy. Creating a device that is both efficient and cost-effective requires materials that are easy to work with and manufacturing methods that are simple and reliable. Since most inorganic state-of-the-art thermoelectric materials lack flexibility, require scarce elements, and are toxic,[1–7] there is a growing desire to explore organic thermoelectric (OTE) materials that, while often less efficient, address these concerns. These OTE materials offer an alternative that is attractive for its flexibility and low thermal conductivity, [8,9] but generally lacks the performance necessary to power a wearable device. Because of the limitations of both inorganic and organic-based materials, an obvious alternative is a tailored polymer-based composite. In our previous work, we found that combining *Polydimethylsiloxane* (PDMS), silica, and carbon nanotubes (CNTs) as a three-material composite created a significant increase in overall device thermopower. This increase was attributed both to the excluded volume effect discussed at length in Park et al.[10], and tunneling transport theory described in the work of Prabhakar et al. [11]While there has been extensive research on the effect of CNTs on polymer-based materials, these studies are often limited to two material composites, and generally focus on reducing the lattice thermal conductivity or increasing power factor by increasing CNT content.[12] In our study we explore and analyze the impact of varying the silica content and doping concentration (both p-type and n-type) on the properties critical for thermoelectric energy conversion. We hope that this study can offer valuable insight into the improvement of PDMS/Silica/CNT composites for their use in wearable and flexible technologies.

As thermoelectric energy conversion devices often employ alternating n-type and p-type semiconductive materials, our study investigates two similar three-part composites. Forming the bulk of both composites is a mixture of PDMS and Silica microparticles. The third part of each composite is comprised of CNTs doped n-type or p-type respectively. We have selected materials based on both their material properties, long-term stability, and for their viability for use in thermoelectric applications. This last criterion covers a number of variables, which are incorporated into the dimensionless figure of merit (zT). Used rigorously through in the study of thermoelectric materials, it considers various properties like the Seebeck coefficient ($\mu V K^{-1}$), electrical conductivity ($S cm^{-1}$), thermal conductivity ($W m^{-1} K^{-1}$), absolute temperature ($K$), and power factor $(PF) = S^2\sigma (\mu W cm^{-1} K^{-1})$. PDMS is an excellent material choice, because of its low thermal conductivity, biocompatibility and density. It exhibits good flexibility across a wide temperature range, and performs well as a binding material.[11–13] We had examined Silica microparticles in our previous studies, and had found that a 3μm particle size optimized composite performance and would create a low-cost filler network, while simultaneously increasing both the electrical conductivity and Seebeck coefficient of the CNT-PDMS based composite. Summarizing the excluded volume effect, described in detail in Park et. al.[10], the large size of the silica particles relative to both the CNT materials and the PDMS polymer chains, forces the smaller materials into narrow areas between the larger ones. This creates pathways within the composite that are dense with CNTs which enhances the conductivity of the overall composite, without adding significant thermal conductivity. While undoped CNTs are relatively stable and exhibit excellent electrical properties, the effect of doping them complicates things significantly and needs to be addressed individually for n-type and p-type materials. For n-type doping of CNTs, long-term stability, due to volatilization and autoxidation, is often the governing material limitation, so we selected 1,5,7-triazabicyclo [4.4.0] dec-5-ene (TBD) which has been observed to be stable over long periods (~5000h) and over a range of temperatures.[14] Similarly, p-type doping treatments that exhibit high conductivities with dopants such as $HNO_3$ and $SOCl_2$ become ineffectual as a consequence of slow desorption at room temperature and rapid at elevated temperatures also inducing hysteresis apart from likely degradation due to exposure to concentrated acid solutions. [12,15,16]. We instead selected a single electron oxidant, triethyloxonium hexachloroantimonate (OA) to offset these issues.[14,17,18] Because the doping distribution is critical to the thermoelectric properties of the final materials and too much or too little dopant concentration can lead to undesirable results [12,18–21], we evaluated several methods of doping and decided that a non-stoichiometric, solution-based process method would be sufficient to homogeneously mix the dopants and would be a good choice for device fabrication due to its process simplicity.

**Material Characterization and Analysis**

The three-part PDMS/Silica/CNT composite samples were prepared with varying PDMS/Silica concentrations and with a fixed 10% CNT concentration by weight. We also changed the OA and TBD doping concentrations (respectively for p-type and n-type doping) on the CNT material and evaluated each doping concentration at each PDMS/Silica ratio to see what effect it would have on material performance. This rigor gave us a total of 45 unique (5 undoped for control and 20 of each p-type and n-type) three-material composites for testing and evaluation. SEM images were obtained for doping levels of 0.5 µg/ml (Figure 1) and 500 µg/ml (Figure 2) respectively to provide important information about the structure and morphology of the composite materials. Though there is some porosity visible in these images, they show the PDMS acting as a binder between silica particles and CNT materials giving all these composite variants their mechanical strength and flexibility. The SEM images also confirm the expected effect of excluded volume, showing that particularly at higher concentrations of silica (Figure 1 d,h, Figure 2 d,h) the CNT materials tend to form bundled networks in between silica particles. For TBD-doped material, these images additionally show even doping distribution and homogeneity of the overall n-type composite at all concentrations. There was some concern that we would see agglomeration of the TBD material based on the attractive forces of these negatively charged materials[22,23], but the SEM images of TBD materials see little to no agglomeration. Conversely, for OA doped material, the p-type composites show good homogeneity of the composite itself, but particularly at higher OA concentrations of both silica and OA, we see precipitation and agglomeration of dopant materials that could cause unexpected or undesirable results. We attribute this agglomeration to our solution processing technique, which was chosen for its simplicity. As the method of doping was kept consistent, the conclusion can be drawn that silica concentration affects the doping process, and that achieving a more uniform dopant distribution may require other more sophisticated doping methods.

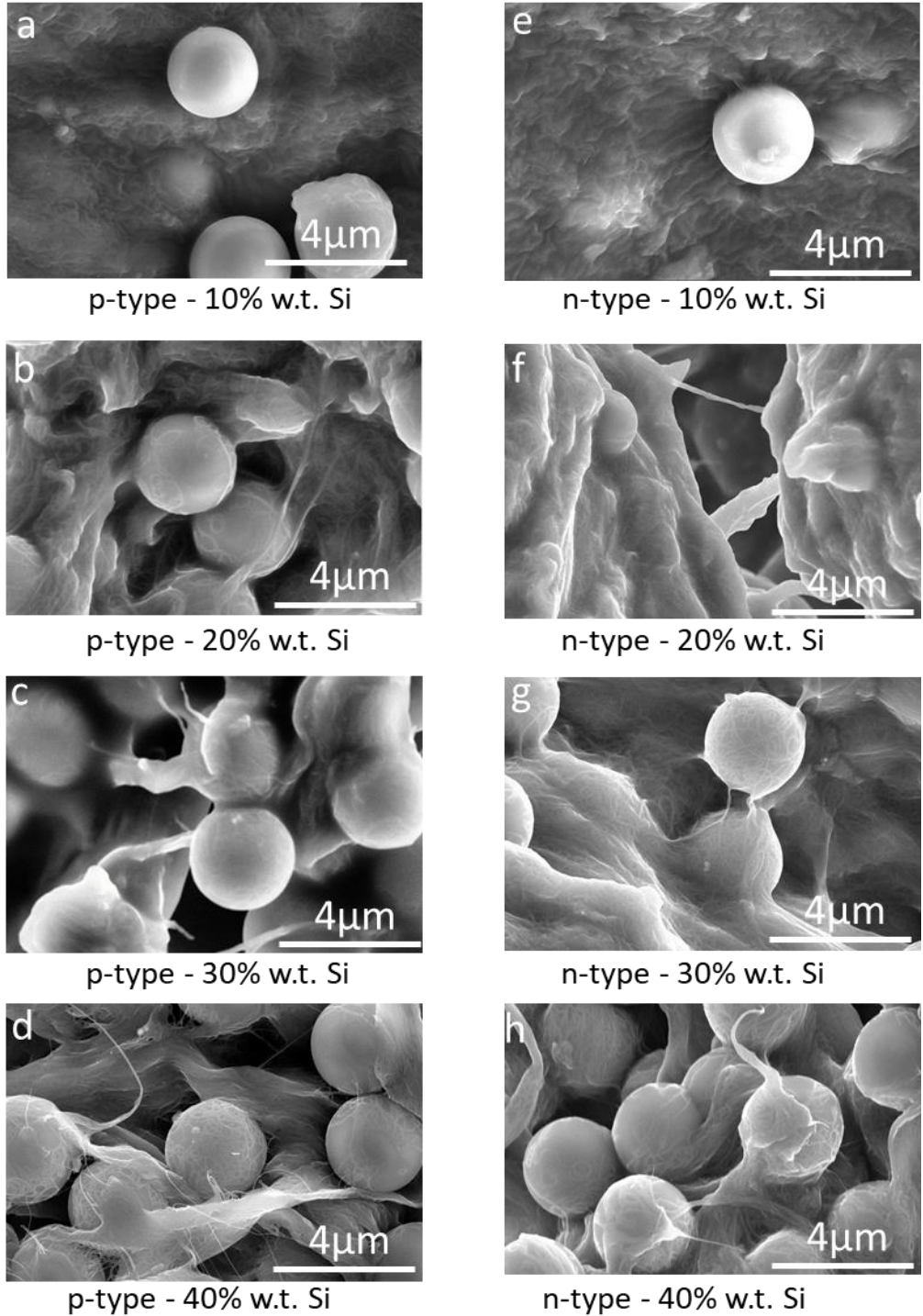

Figure 1: SEM images for carbon-based polymer composites, silica size=3μm with both OA and TBD doping concentration of 0.5 μg/ml

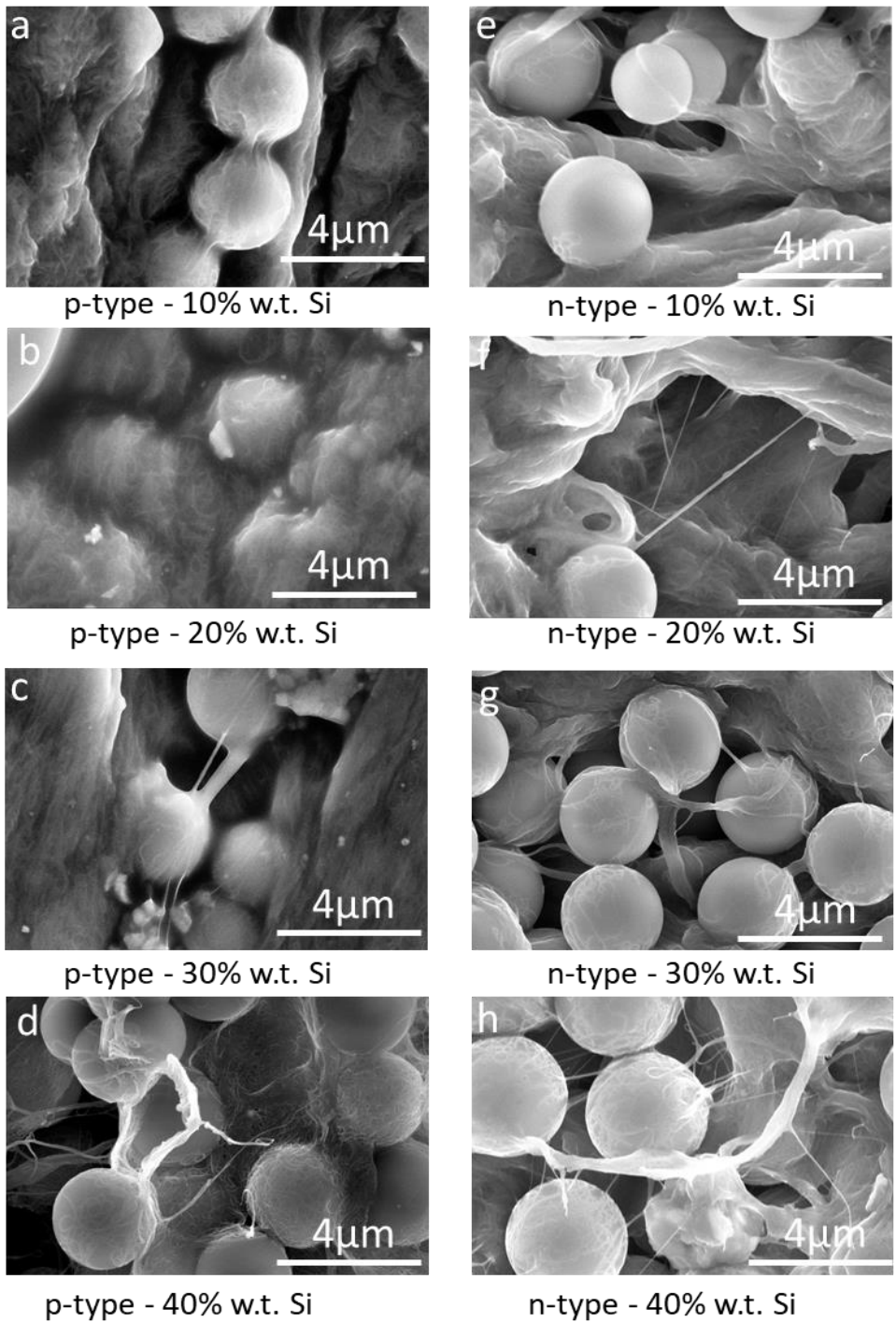

Figure 2: SEM images for carbon-based polymer composites, silica size=3μm with both OA and TBD doping concentration of 500 μg/ml

**Raman Spectroscopy**

Raman spectroscopy is a powerful tool for the characterization of the molecular and chemical structure of a material. In the case of the doped CNT-PDMS-Silica composite with different concentrations of OA, Raman spectroscopy is used to study the effect of doping on the D and G band intensity ratios and the overall band intensity. The D and G bands in Raman spectra of carbon nanotubes are related to the structural defects and curvature of the tubes, respectively. Increasing the concentration of OA doping can introduce new defects and change the curvature of the CNTs, which can affect the D:G ratio. Additionally, the overall band intensity of the composite can be influenced by the doping concentration, as the presence of OA can change the mechanical and electronic properties of the CNTs. In this study, Raman spectroscopy was performed on a Renishaw InVia Raman Microscope using a 633nm laser at a power level of 0.5% and an acquisition time of 10s. The samples were characterized by comparison of the graphitic (G; 1590 cm$^{-1}$) and disordered (D; 1320 cm$^{-1}$) peak areas (Figure 3). The D:G ratio decreased as the doping level increased, confirming chemical interactions between OA and the CNT material. The control sample with no p-doping showed a D:G ratio of 0.0165, the sample prepared using 0.5 µg/ml had a D:G ratio of 0.015, and the sample prepared using 500 µg/ml had a D:G ratio of 0.0098. This trend suggests that while OA introduces charge carriers as holes by oxidation of the sample, the graphitic character increases, suggesting interaction of the doping agent with the sample.

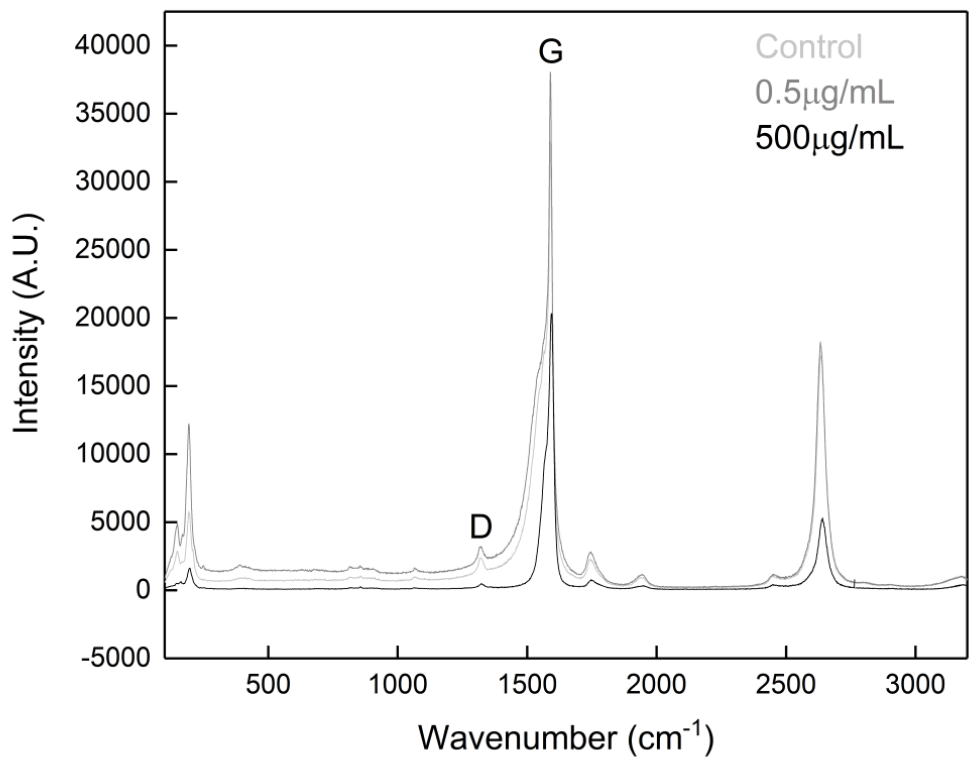

Figure 3: Raman spectra of the unmodified control sample, a sample doped with OA at 0.5 μg/mL, and a sample doped with OA at 500 μg/mL. The D:G ratio was calculated based on the areas of the peaks at 1320cm$^{-1}$ (D) and 1590cm$^{-1}$ (G). As the concentration of OA in the sample increased, the D:G ratio decreased implying an increase in the ratio of disordered carbon to graphitic carbon. This means that the samples underwent a slight increase in graphitic character as the doping concentration increased.

FTIR-ATR was performed on the PDMS/Silica/CNT composite samples. The sample spectra were compared to pure PDMS and pure silica. The most prominent peaks in the samples represented a combination of the pure PDMS and silica, with peak positions at ~3000cm$^{-1}$, 1260cm$^{-1}$, 1070 and 1015cm$^{-1}$, and ~850cm$^{-1}$ assigned as C-H, Si-CH$_3$, Si-O-Si, and Si-CH$_3$, respectively. The carbon nanotubes themselves consist primarily of a sp$^2$-hybridized carbon network, the bonds of which will not be detected by this technique as IR absorption requires bonds with an existing dipole moment.

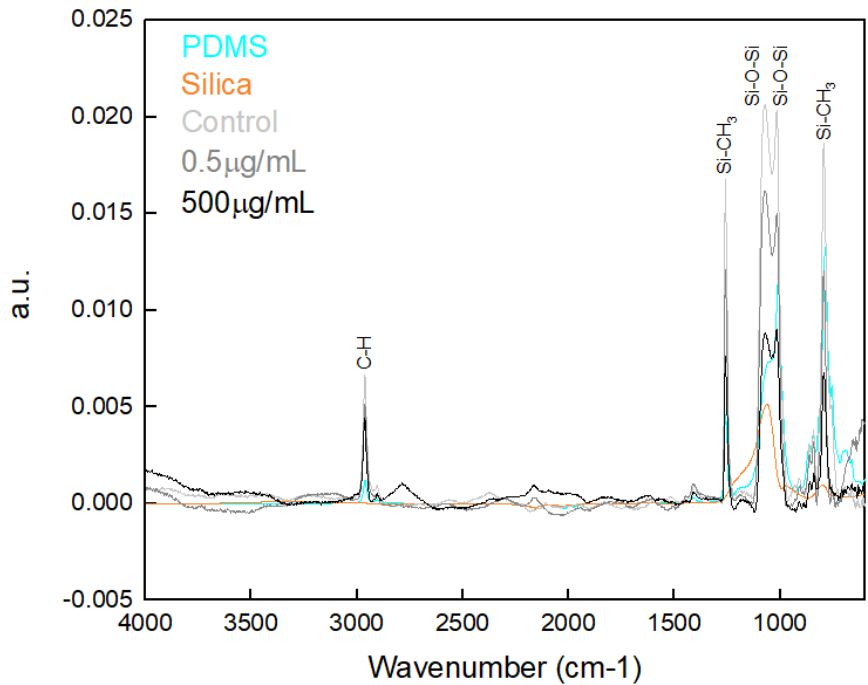

Figure 4: FTIR-ATR Spectra of (blue) pure PDMS, (brown) pure silica, (light grey) control sample, (dark grey) sample doped with 0.5μg/mL OA and (black) sample doped with 500μg/mL OA. All samples showed a combination of the characteristic peaks expected from the corresponding samples: C-H at ~3000cm$^{-1}$, Si-CH3 at ~1260cm$^{-1}$, Si-O-Si at ~1015 and ~1070cm$^{-1}$ and Si-CH3 at ~850cm$^{-1}$ in the pure PDMS and all CNT samples, and the aforementioned Si-O-Si peak in the pure silica sample.

**Thermoelectric Characteristics of the Doped PDMS/Silica/CNT Composite**

In an evaluation of the doped PDMS/Silica/CNT composite, it is demonstrated that the Seebeck coefficient and electrical conductivity were affected by both type and concentration of the dopant used, and by the amount of silica present in the sample. In the case of n-type doping with TBD, increasing the silica content increases both the Seebeck coefficient (in magnitude) and electrical conductivity. This trend occurs at every concentration of TBD dopant used (Figure 5b and 6b respectively). Electrical conductivity follows a linear trend, while the Seebeck coefficient follows an exponential one. This is attributed to the excluded volume effect, where microscale silica particles channel the smaller doped CNT materials between them. This is highly advantageous to thermoelectric materials because of the contribution to zT made by the Seebeck coefficient. Also notable was the effect of TBD doping concentration on the Seebeck coefficient. Though higher TBD dopant concentrations show better electrical conductivity at every Silica concentration, the Seebeck coefficient shows the opposite. The lowest doping concentration not only provided the best Seebeck coefficient but also showed a faster exponential curve as silica content increased.

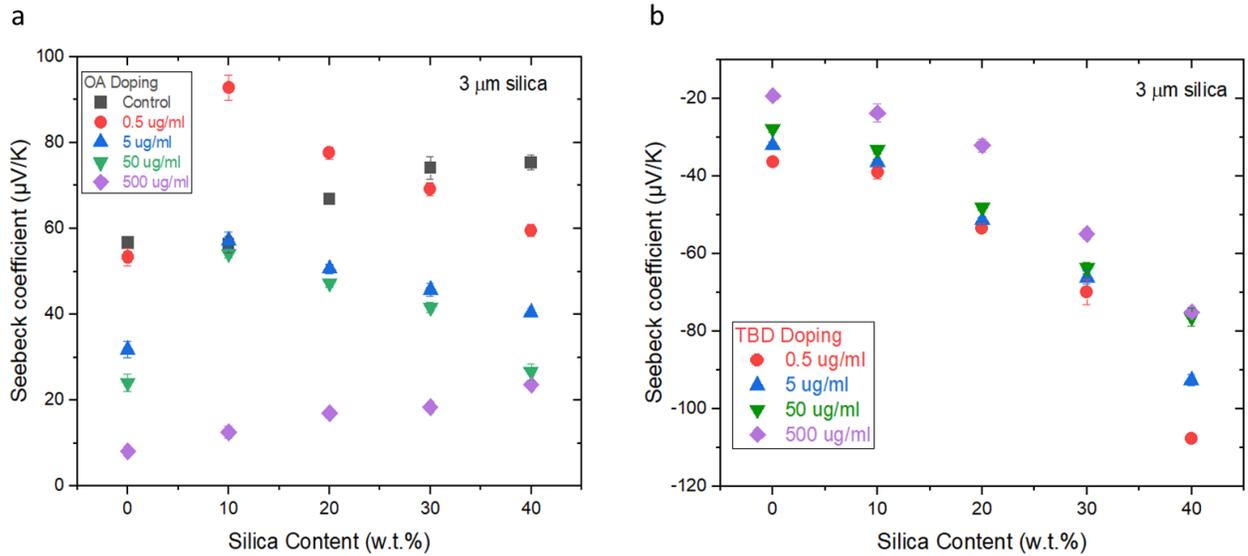

Figure 5: (a) Seebeck coefficient of OA doped SWCNT-silica MP-PDMS composites as a function of silica MP w.t.% for 3 μm particle size (b) Seebeck coefficient of TBD doped SWCNT-silica MP-PDMS composites as a function of silica MP w.t.% for 3 μm particle size

Our best n-type material observed was 40% wt. silica and 0.5μg/ml TBD which maximized the power factor at 28.46 $\mu W/mK^2$ (Figure 7b). Knowing that long-term stability is often a concern for n-type materials, we tested this material after 45 days, measuring both Seebeck and electrical conductivity (Figure 8). Over that time period, we recorded a drop of roughly 18% in both the Seebeck coefficient and electrical conductivity. While this decay was substantial, our goal was to review the composite without alteration or encapsulation, so there were no steps taken outside of material selection to extend long-term stability. So, while the results of this material are quite promising, and exhibit some resistance to decay, it is not a final solution, and steps would need to be taken to encapsulate the TBD material before it could be used commercially. With those caveats, the implication here is for n-type doping, we can achieve excellent thermoelectric properties while using only a fraction of the CNT materials. By comparison, the OA-doped CNT material had a much less uniform result. At low dopant concentrations, adding silica first created a spike in Seebeck, and a reduction in electrical conductivity.

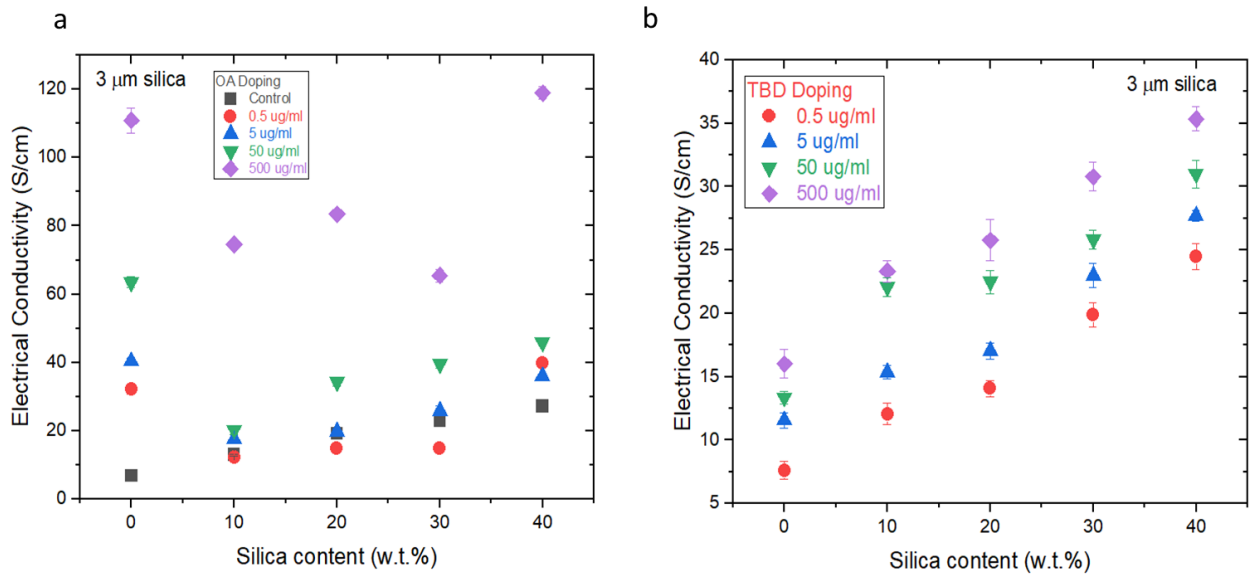

Figure 6: (a) Electrical conductivity of OA doped SWCNT-silica MP-PDMS composites as a function of silica MP w.t.% for 3 μm particle size (b) Electrical conductivity of TBD doped SWCNT-silica MP-PDMS composites as a function of silica MP w.t.% for 3 μm particle size

As more silica was added, the Seebeck coefficient went down, while electrical conductivity increased. Conversely, at high OA doping concentrations, both the Seebeck and electrical conductivity increase slowly with increased silica content, and while electrical conductivity reaches the highest value demonstrated by any of the p-type composites, the Seebeck coefficient is still lower than in any of the other materials. It is understood that with an increase in the amount of OA adsorbed, the carrier density correspondingly increases causing the Fermi level ($E_f$) to move away from the center of the energy gap. This explains a systematic increase in conductivity and a decrease in the Seebeck coefficient at the same time.[12] Ultimately what was observed was that low OA concentrations combined with low concentrations of silica made the best p-type materials of those we studied, but that these still exhibited lackluster performance compared to their n-type counterparts. We selected 10% wt. silica and 0.5µg/ml OA as our p-type material for device study even though it only exhibited a power factor of 10.41 µW/mK².

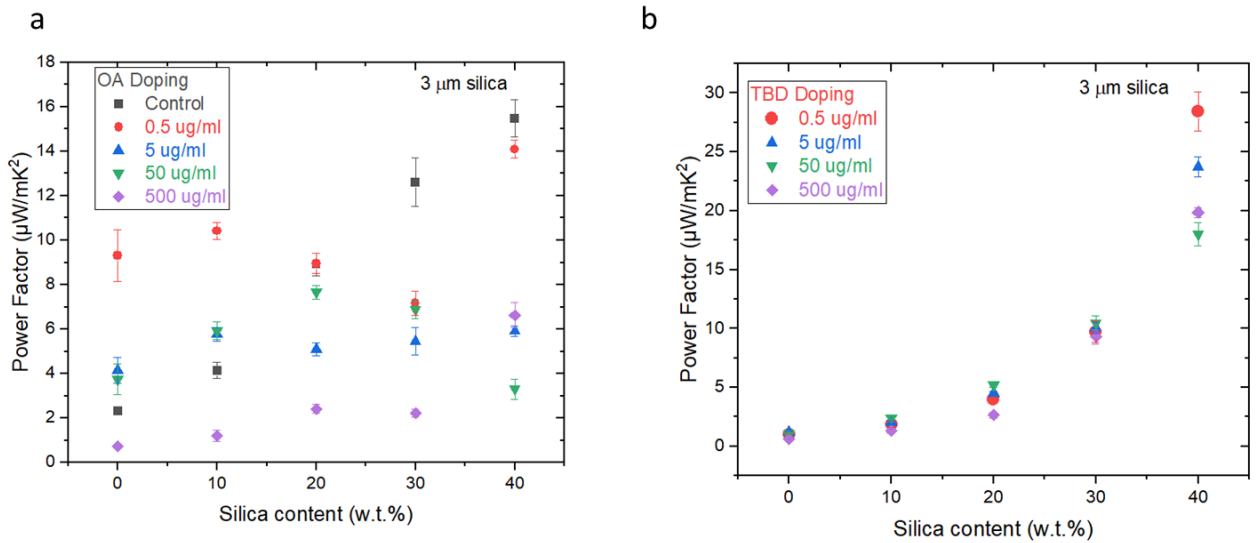

Figure 7: (a) Power factor of OA doped SWCNT-silica MP-PDMS composites as a function of silica MP w.t.% for 3 µm particle size (b) Power factor of TBD doped SWCNT-silica MP-PDMS composites as a function of silica MP w.t.% for 3 µm particle size

In comparison to other studies involving CNT and PDMS, we found that the effect of adding silica is quite substantial. Utilizing a fixed 10% undoped CNT in our control sample while altering silica concentration gives us a fairly one-to-one comparison to the study performed by Prabhakar et. al. [11] which reports a power factor of 18 µW/mK² at 40% CNT. In our case, we see a steady increase in power factor with increased silica content and achieve a comparable power factor of 14.09 µW/mK² at 40% silica using only one-fourth of the CNT material. This observation is made with the concession to material properties (i.e. reduced flexibility and stretchability) that adding silica inherently causes. With our prior study working with these materials, we also did not pursue increasing silica content beyond 40% as it both negatively impacts properties, and has a significant negative impact on thermoelectric properties.

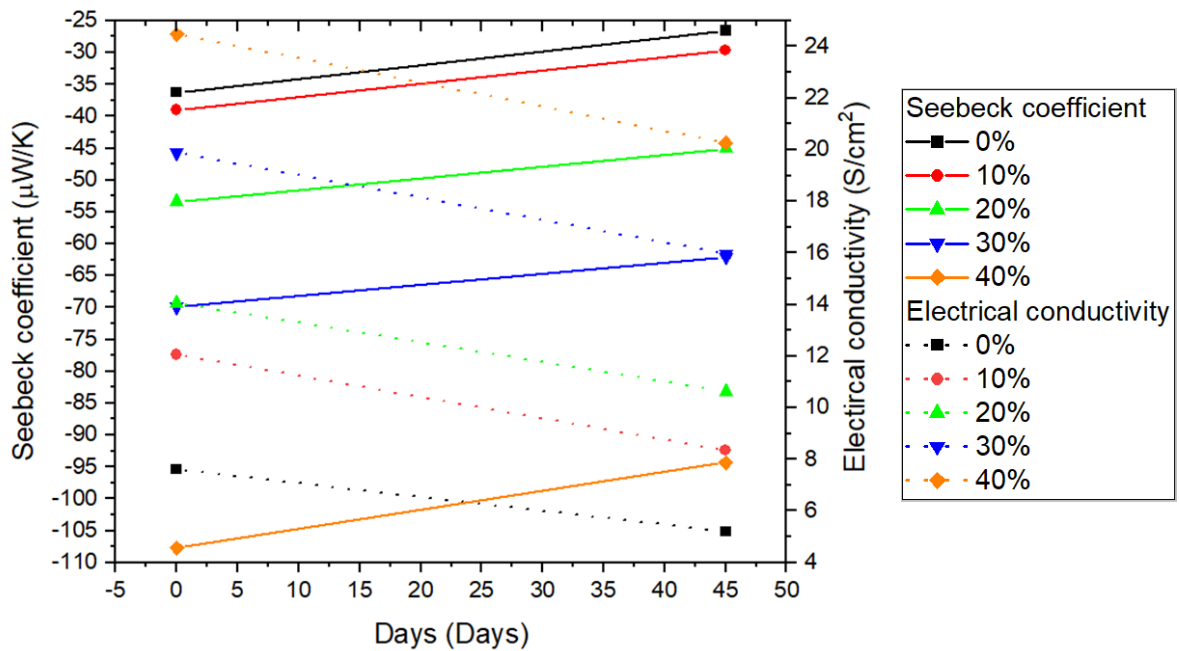

Figure 8: Stability test of both the Seebeck and electrical conductivity of TBD doped samples

## Device Fabrication and Testing

After a thorough analysis of the composite materials we had, we went on to apply this knowledge to the fabrication of a device. To demonstrate the TE voltage output and power production of these PDMS/Silica/CNT composites, a flexible TE generator composed of 27 pairs of p-n junctions coupled in series was built into a relatively lightweight (23.6g) flexible band (Figure 8 a,b). With overall dimensions of 230mm X 20mm X 5mm, this TE generator produced a maximum power output of 1.15 µW (0.025 µW/cm²) with a hot side temperature of 36°C (average skin temperature), and a cold side temperature of 32°C, (ΔT = 4K). The molded device strap was created from preformed PDMS, chosen for its softness, lack of toxicity, flexibility, and low thermal conductivity [24].

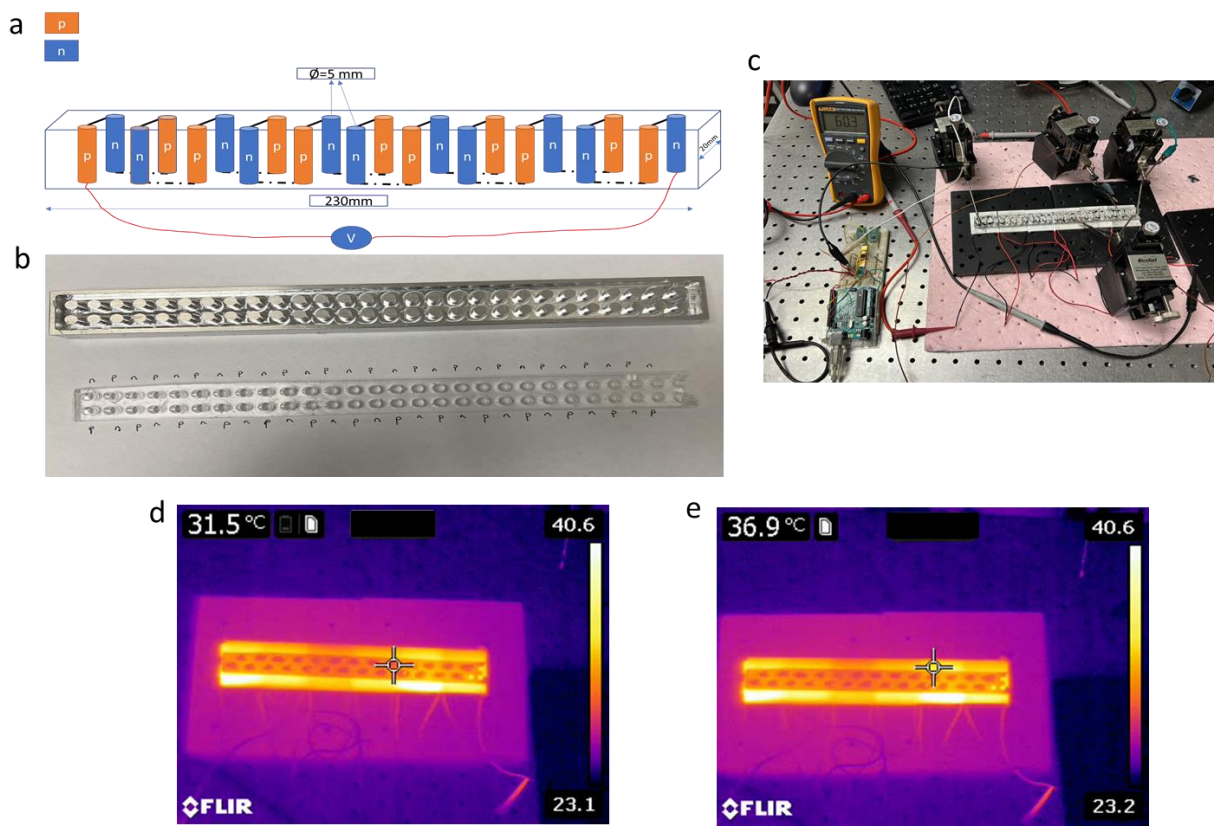

Figure 9: (a) A schematic illustration of the number of p-n leg pairs used to construct the TE device. (b) An aluminum mold utilized to cast the PDMS flexible band. (c) A benchtop setup employed to connect the modules and produce the necessary current. (d) and (e) Images demonstrating the temperature gradient using a FLIR spot infrared thermometer.

Based on their thermoelectric properties, we selected 40% w.t. silica, n-type $0.5 \mu g/ml$ and 10% w.t. silica p-type $0.5\ \mu g/ml$ from our list of available composites. These were then extruded into a conventional $\pi$-type configuration (Figure 9a) and connected electrically in series using silver paste.

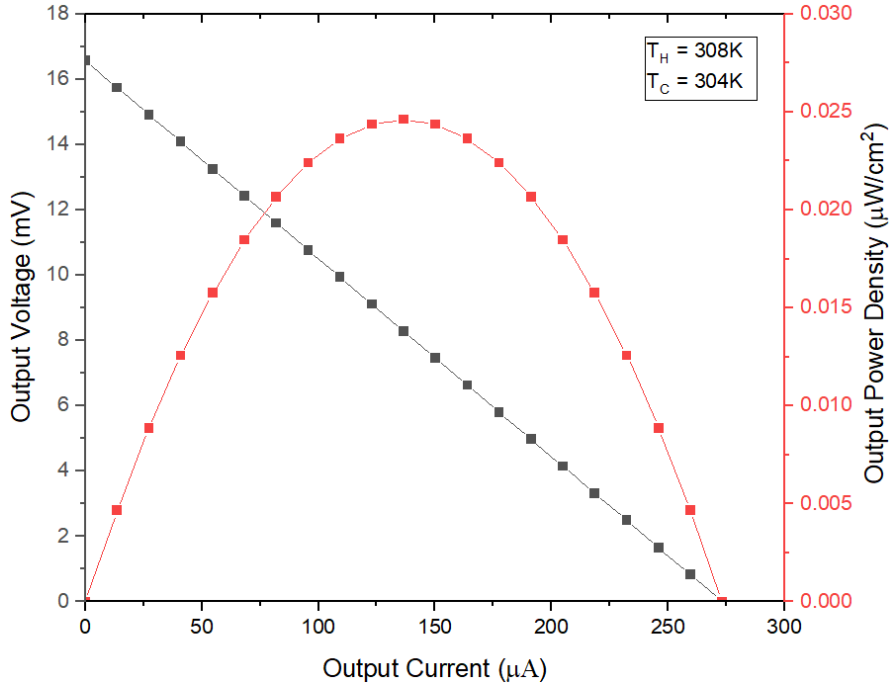

Figure 10: Temperature and output voltage as a function of heating time for TE device

The measurements obtained show that this TEG can generate an open-circuit voltage of 1.385 m V at $\Delta T = 4\ K$ and the internal resistance of the device was recorded to be approximately 60 Ω across a working temperature suitable to wearable devices. In the benchtop analysis, a hot side steady state temperature of 36 °C (308 K) was maintained to emulate skin temperature, whereas the cold side was exposed to ambient air for a steady heat flux in series with a matched load[25]. The device exhibits a considerable maximum power output of 1.15 µW (0.025 µW/cm$^2$) at $\Delta T = 4\ K$. The result is device suitable for a number of wearable applications, as it considers factors such as internal resistance, flexibility, and form factor. To highlight the effectiveness of a doped

PDMS/Silica/CNT composite, we look at past studies of flexible thermoelectric devices, Yang et. al. [26] and Shi et. al [27] reported inorganic-based ($Bi_2Te_3$) wearable thermoelectric generators (TEGs) using flexible printed circuit board (FPCB) that generated 130.6 nW at a 12 K temperature difference, and 23 µW at a temperature difference of 35K respectively. In a related application, Lee et. al. discusses TEG based on soft heat conductors (s-HCs) and intrinsically stretchable interconnects with 220-np-pairs exhibiting a maximum power of 7.02 mW with temperature difference of 40K [28].

**Conclusion**

While this study presents a rigorous examination of 45 unique composites (5 control and 20 of each p-type and n-type) and our results both confirm anticipated trends and produce excellent results, it is clear that there is an ocean of exploration ahead in terms of organic composite materials for thermoelectric use. The resultant p-type (OA triethyloxonium hexachloroantimonate) composites underperformed by comparison to the n-type counterparts (TBD 1,5,7-triazabicyclo [4.4.0]dec-5-ene), but their evaluation in this study makes an excellent point of comparison. It demonstrates some of the obstacles to overcome by utilizing these techniques and gives a great opportunity for improving future device performance. Though we achieved a maximum power factor of 28.46 µW/mK$^2$ with our TBD material, and an overall device output power of 1.15 µW (0.025 µW/cm$^2$) at a temperature difference of only ΔT = 4 K, we recognize how many variables we left unexamined, and how much optimization still could be done.  We acknowledge that our TBD material achieving only an 18% drop in performance over 45 days (Seebeck and conductivity) may be an excellent result today, but that it leaves a lot of room for stability improvement for future processes, and for use as a commercially viable material. We are excited to report these findings on PDMS/Silica/CNT composites as this level of power output is adequate for a broad range of wearable devices, and we hope this study provides a springboard for future analysis of other three multi-material composites.

**ACKNOWLEDGMENT**

This material is based upon work supported by the National Science Foundation under Grants No. 1905571 and 1905037. V.S. and V.K.R.K. thank support from NIOSH through a grant #T42OH008432 from the Pilot Research Project Training Program of the University of Cincinnati (UC) Education and Research Center, NSF [grants IIP-2016484 and CBET-2028625], and the UC Collaborative Research Advancement Grant No. 1018371.